\documentclass[prl,10pt,floatfix,twocolumn,aps]{revtex4-1}
\setcitestyle{super}
\usepackage{hyperref}
\usepackage{graphicx}
\usepackage{amsmath}
\usepackage{color}
\usepackage{amsfonts}
\newcommand*{\citen}[1]{%
  \begingroup
    \romannumeral-`\x 
    \setcitestyle{numbers}%
    \cite{#1}%
  \endgroup   
}

\newcommand{\dois}[2]{\href{http://dx.doi.org/#1}{#2}}

\begin{document}

\title{Spiral bandwidth of four-wave mixing in Rb vapour}

\author{R.\ F.\ Offer,$^1$ D.\ Stulga,$^1$ E.\ Riis,$^1$ S.\ Franke-Arnold,$^2$ and A.\ S.\ Arnold$^{1*}$}

\affiliation{$^1$Department of Physics, SUPA, University of Strathclyde, Glasgow G4 0NG, UK\\
$^2$School of Physics and Astronomy, SUPA, University of Glasgow, Glasgow G12 8QQ, UK\\$^*$Corresponding author: aidan.arnold@strath.ac.uk}





\maketitle

\noindent {\large {\bf \textsf{Abstract}}}
\\ 
\textbf{Laguerre-Gauss beams, and more generally the orbital angular momentum of light (OAM) provide valuable research tools for optical manipulation, processing, imaging and communication. High-efficiency frequency conversion of OAM is possible via four-wave mixing in rubidium vapour. Conservation of the OAM in the two pump beams determines the total OAM shared by the generated light fields at 420$\,$nm and 5.2$\,\mu$m -- but not its distribution between them. Here we experimentally investigate the spiral bandwidth of the generated light modes as a function of pump OAM.  A small pump OAM is transferred almost completely to the 420$\,$nm beam.  Increasing the total pump OAM broadens the OAM spectrum of the generated light, indicating OAM entanglement between the generated light fields. This clears the path to high-efficiency OAM entanglement between widely disparate wavelengths.} \\


\noindent {\large {\bf \textsf{Introduction}}}\\
The orbital angular momentum (OAM) of light\cite{Franke-Arnold2008}, electrons\cite{barnett} and neutrons\cite{clark} is a burgeoning research topic. OAM and the transverse profile of light more generally, have a myriad of uses including the optical tweezing and spinning of dielectric particles, micro-machines and biological specimens, as well as for guiding and rotating ultracold atoms\cite{Franke-Arnold2008,Franke-arnold2017}. Moreover, light's OAM has the potential to greatly increase the bandwidth of both classical\cite{Wang2012} and quantum communication \cite{Langford2004,Groblacher2006,Mafu2013}.  In order to fully realise this potential, methods of frequency converting, manipulating and generating classical and quantum OAM states are required.  

Phase-matched nonlinear processes provide the means to undertake many of these operations.  The key is that these processes are phase coherent -- both longitudinally and transversely -- for the pump beams and generated light.  As a consequence OAM, which is associated with spiral phase fronts \cite{Franke-arnold2017}, must be conserved \cite{Dholakia1996,Mair2001}, and more generally nonlinear processes can be used to manipulate  transverse light modes.  Frequency conversion of transverse modes \cite{Steinlechner2015}, images \cite{Ding2012} and entangled OAM states \cite{Zhou2016} has been demonstrated via sum-frequency generation, whilst OAM addition has been carried out in nonlinear crystals \cite{Dholakia1996,Li2015} and atomic vapours \cite{Walker2012,Akulshin2016,Pruvost2018}.

When a phase-coherent process generates two photons they are highly correlated -- phase matching ensures that e.g.\ the total momenta and total orbital angular momenta of the outgoing photons match that of the pump photon(s).  Spontaneous parametric down conversion (SPDC) is routinely used to produce OAM-entangled photon pairs \cite{Mair2001,Franke-Arnold2002}, four-wave mixing (FWM) in atomic vapours has been used to both create \cite{Boyer2008} and store \cite{Marino2009} entangled images, and FWM in cold atoms has very recently been shown to produce photon-atom entanglement between the OAM of the pump photon and the atomic spin wave, and in consequence between the generated photons \cite{Shi2018}.    
Correlations are also observed in properties other than the transverse mode.  Intensity-difference squeezed beams have been generated via FWM in atomic vapours \cite{Ma2017}, and intensity correlations have been transferred from one field to another \cite{Ihn2017}.  Similar systems have been used to prepare heralded bichromatic single photons \cite{Whiting2017}, and polarisation entangled photons have been produced with cold atoms \cite{Chaneliere2006}.

In this work we quantitatively explore the transfer of OAM between fields involved in a resonantly enhanced FWM process in rubidium vapour. Two near-infrared pump fields (780$\,$nm and 776$\,$nm) generate an infrared and a blue field (5.2$\,\mu$m and 420$\,$nm) \cite {Zibrov2002,Meijer2006,Akulshin2009,Vernier2010}.
To study OAM transfer in FWM we use Laguerre-Gauss (LG) pump modes, which are characterised by their azimuthal and radial indices, $\ell\in\mathbb{Z}$ and $p\in\mathbb{Z}^*$, with each mode carrying $\ell\hbar$ of OAM \cite{Allen1992} per photon and a relative electric field in cylindrical polar co-ordinates $(r,\theta,z)$ given by:
\begin{equation} 
\textrm{LG}^\ell_p=\frac{C^\ell_p}{w}\left(\frac{r\sqrt{2}}{w}\right)^{\!\!\vert \ell\vert} \!\! L^{\vert \ell\vert}_p\!\!\left[\frac{2r^2}{w^2}\right] \, e^{-\frac{r^2}{w^2}+i (k z+\ell \theta+\Phi)},
\label{eq:LG field}
\end{equation}
where ${C^\ell_p=\sqrt{2p!/\pi(p+\vert \ell\vert)!}}$, $L^{\vert \ell\vert}_p$ is an associated Laguerre polynomial, $w=w_0 (1+\left(z/z_\textrm{R}\right)^2)^{1/2}$ is the beam $e^{-2}$ radius for a waist $w_0$, $z_\textrm{R}=\pi w^2_0/\lambda$ is the Rayleigh range, and $\Phi=\Phi_\textrm{S}+\Phi_\textrm{G}$ is the sum of the spherical phasefronts $\Phi_\textrm{S}=k r^2 z/(2(z^2+z_\textrm{R}^2))$ and Gouy phase $\Phi_\textrm{G}=-(2p+\vert \ell\vert+1)\arctan(z/z_\textrm{R})$. 

Conservation of OAM requires that the generated fields with infra-red $\lambda_\textrm{IR}=5.2\,\mu$m and blue $\lambda_\textrm{B}=420\,$nm  wavelengths, and associated OAM indices $\ell_\textrm{IR}$ and $\ell_\textrm{B},$ must have the same total OAM $\ell_\textrm{T}$ as supplied by the pump fields. However, the pump OAM can be shared in different ways between the two generated fields as long as: $\ell_\textrm{T}=\ell_{780}+\ell_{776}=\ell_\textrm{IR}+\ell_\textrm{B}$, i.e.\ angular phase matching occurs. Although currently the 5.2$\,\mu$m field is absorbed by the glass of our cell, it could be observed from sapphire cells \cite{Zibrov2002,Akulshin2014a}. Based on recent experiments using four-wave mixing in a cold Rb vapour, albeit with a more degenerate level scheme \cite{Shi2018}, we infer that at the quantum level we will efficiently generate an entangled output photon state of the form: 
\begin{equation}
|\Psi\rangle_{\ell_\textrm{T}} = \sum_{\ell_\textrm{B}, p_\textrm{B}, p_\textrm{IR}} c_{p_\textrm{B},p_\textrm{IR}}^{\ell_\textrm{B},\ell_\textrm{IR}} |\textrm{LG}_{p_\textrm{B}}^{\ell_\textrm{B}}\rangle_\textrm{B}\;
|\textrm{LG}_{p_\textrm{IR}}^{\ell_\textrm{IR}}\rangle_\textrm{IR}, 
\label{entangledstate}
\end{equation}
with the constraint $\ell_\textrm{IR}=\ell_\textrm{T}-\ell_\textrm{B}.$
This allows us to infer the spiral bandwidth\cite{Torres2003} or $\ell$-distribution standard deviation of:
\begin{equation}
\Delta \ell(\ell_\textrm{T}) = 
\sqrt{\sum_{\ell_\textrm{B}}
{ P_{\ell_\textrm{B},\ell_\textrm{IR}}
 \; {\ell_\textrm{B}}^2}-\left(\sum_{\ell_\textrm{B}}
{ P_{\ell_\textrm{B},\ell_\textrm{IR}}
 \; \ell_\textrm{B}} \right)^2}\; ,
\label{spiralbandwidth}
\end{equation}
which is a measure of how many orthogonal modes could be entangled in such a state. The associated Shannon entanglement entropy (information content) \cite{shan} is
\begin{equation}
S(\ell_\textrm{T}) = - \sum_{\ell_\textrm{B}=-\infty}^{\infty} P_{\ell_\textrm{B},\ell_\textrm{IR}} \log_2 P_{\ell_\textrm{B},\ell_\textrm{IR}} \; ,
\label{entanglemententropy}
\end{equation}
where $P_{\ell_\textrm{B},\ell_\textrm{IR}}=\sum_{p_\textrm{B}, p_\textrm{IR}}{|c_{p_\textrm{B},p_\textrm{IR}}^{\ell_\textrm{B},\ell_\textrm{IR}}|^2}$ is the total probability the output photon pair is in a state of the form $|\textrm{LG}_{p_\textrm{B}}^{\ell_\textrm{B}}\rangle_\textrm{B}\;
|\textrm{LG}_{p_\textrm{IR}}^{\ell_\textrm{T}-\ell_\textrm{B}}\rangle_\textrm{IR}.$ We do not currently measure at the single photon level, but can deduce the coefficients $c_{p_\textrm{B},p_\textrm{IR}}^{\ell_\textrm{B},\ell_\textrm{IR}}$ in Eq.~\ref{entangledstate} from a detailed Fourier analysis of the generated blue light beam and hence evaluate the spectral bandwidth (Eq.~\ref{spiralbandwidth}) and entanglement entropy (Eq.~\ref{entanglemententropy}). 

Many methods to quantitatively measure the OAM spectrum of a beam or even a single photon have been demonstrated, including fork diffraction gratings used as filters followed by on-axis detection \cite{Mair2001,Flamm2012,Forbes2016,Rui2016}, cascading Mach-Zehnder interferometers containing Dove prisms \cite{Leach2002,Gonzalez2006} and transformation optics \cite{Berkhout2010,Dudley2013}.  All methods are applicable to both coherent and incoherent superpositions of OAM ($\ell$) modes, however the identification of $p$ modes proves more challenging.  The latter two methods give no or only indirect information on $p$ modes.  Cycling through the relevant fork gratings does, in principle, allow the identification of the complete mode decomposition, but is time-consuming and requires an additional SLM.
Previous vapor FWM experiments have determined the OAM state of the 420$\,$nm emission by visual inspection of either an interferogram\cite{Walker2012} or the beam pattern after a tilted lens
\cite{Akulshin2016,Pruvost2018}.  

Here we perform a full Fourier analysis of the experimental interferograms to precisely identify the modal superposition.  We show that our experimental observations agree with a model\cite{Walker2012,Lanning2017} that incorporates a full ($\ell$ and $p$) LG$_p^{\ell}$ modal description of the four fields. In agreement with past observations \cite{Walker2012,Akulshin2016,Pruvost2018} we find that for low pump OAM, the 420$\,$nm light is generated in an almost pure mode, with an OAM given by the sum $\ell_\textrm{T}$ of the pump modes. 
As the pump OAM increases ($\ell \geq 4$) we observe the 420$\,$nm light in an incoherent superposition of an increasing number of OAM modes, as predicted in Ref.~\citen{Lanning2017}, indicating that the 5.2$\,\mu$m and 420$\,$nm two-photon state becomes OAM entangled.  We obtain the experimental spiral bandwidth and entanglement entropy via the full $\ell$ and $p$-mode decomposition of the generated blue light for a range of pump $\ell$.  We find that both the spiral bandwidth and entanglement entropy increase with pump OAM, but they also depend on how the input OAM is divided between the two pump beams.
This versatility is due to the very different wavelengths of the two generated fields (5.2$\,\mu$m and 420$\,$nm), which leads to the pump OAM being shared between these fields in a way that critically depends on the pump mode. Our results indicate that this system will be an efficient source of OAM entangled photon pairs \cite{Becerra2008,Srivathsan2013,Whiting2018}, as well as a means to add and frequency convert OAM states. \\

\begin{figure}[!b]
\centering
\includegraphics[width=\linewidth]{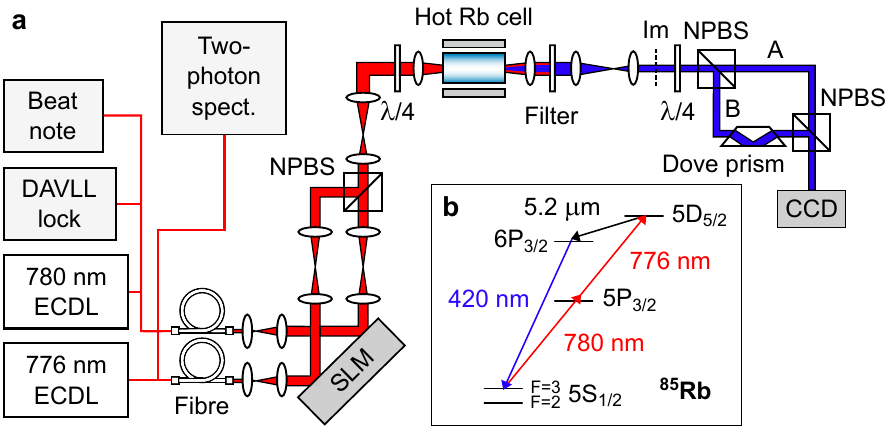}
\caption{The experimental setup. Abbreviations used in the setup (a) are ECDL (extended cavity diode laser), DAVLL (dichroic atomic vapour laser lock), SLM (spatial light modulator), NPBS (non-polarising beam splitter) and CCD (charge-coupled device). (b) The relevant $^{85}$Rb energy levels.}
\label{fig:fig 1}
\end{figure}

\noindent {\large {\bf \textsf{Results}}}
\\   
Details of the experimental setup (Fig.~\ref{fig:fig 1}) are given in the Methods and Supplementary Note 1. Details of the generated light LG$_p^{\ell}$ decomposition are provided in the Methods and Supplementary Notes 2-4.  
\\

\noindent\textbf{Mode decomposition via Fourier analysis.} 

We determine the full incoherent $\ell$ and $p$-mode decomposition of the light profile of interest via Fourier analysis of the simple interferogram formed when the light is overlapped with its mirror image, as can be obtained at the output of a Dove prism interferometer (Fig.~\ref{fig:fig 1}a).  This technique distinguishes our approach from previous experiments with this FWM system \cite{Walker2012,Akulshin2016}, where the 420$\,$nm mode was determined by visual inspection of interferograms\cite{Walker2012} or fringes after a tilted lens\cite{Akulshin2016,Pruvost2018}.   

For a pure LG$_p^{\ell}$ mode, the interferogram has $2\ell$ azimuthal lobes \cite{ferris,ferris2}.  If the FWM light was generated in a coherent superposition of modes, the generated complex light fields would add, leading to azimuthal\cite{ferris,ferris2} or radial interference\cite{ferris2} in the raw beam intensity prior to the interferometer. However, neither our input nor the generated raw light beams show such interference effects and we always observe clean single rings of light. 
This confirms that the light is generated in an incoherent superposition of modes, so the intensities of the constituent modes add, without interference contributions from their cross terms. 

The interferogram is therefore simply the sum of the `ferris wheel' interference patterns of the constituent modes: 
\begin{equation}
I(r,\theta)=\sum\limits_{\ell,p} P^{\ell}_pR^{\ell}_p(r) \left[1+\cos\left(2\ell\theta+\phi^{ \ell}_p\right)\right]/2,
\label{eq:intIncoh}
\end{equation}
where $ P^{ \ell}_p$ is the relative power in each mode, $R^{ \ell}_p$ is the radial intensity profile of each mode and $\phi^{ \ell}_p$ is the mode-dependent interferometer phase.  Note for $\ell=0$ modes the total interferogram amplitude entirely depends on interferometer phase, $\phi^{0}_p,$ and thus the relative power in $\ell=0$ modes cannot be found from the interferogram.

\begin{figure}[!b]
\centering
\includegraphics[width=\linewidth]{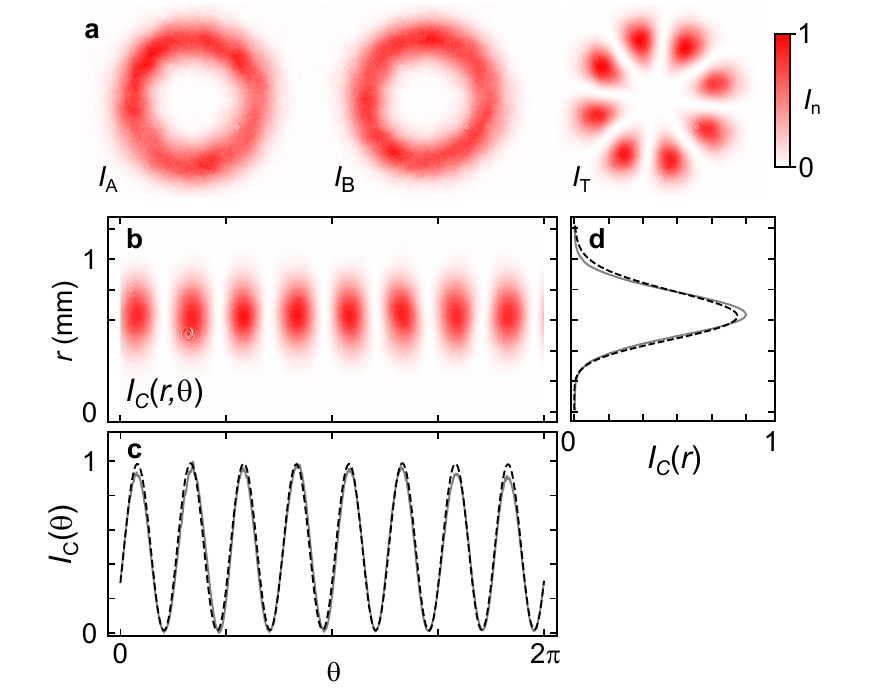}
\caption{Intensity and interferogram of an LG$^{4}_0$ 776$\,$nm pump mode.  (a) Intensity profile in interferometer arm A $(I_\textrm{A}),$ arm B $(I_\textrm{B})$, and the uncorrected interferogram $(I_\textrm{T})$. (b) Corrected interferogram (Eq.~\ref{eq:correction}). (c) Azimuthal and (d) radial profile of the corrected interferogram (solid, gray lines), and reconstructed profiles from the mode decomposition (dashed, black lines). }
\label{fig:fig 2}
\end{figure}

To allow $\ell=0$ mode measurement, and correct for slight discrepancies in 50:50 beam splitter transmission, we take three images at the interferometer output: the interferogram, $I_\textrm{T}$, as well as the intensity profiles in arms A and B, $I_\textrm{A}$ and $I_\textrm{B}$ (Fig.~\ref{fig:fig 2}a).  From theses images we calculate a corrected interferogram $I_\textrm{C}(r,\theta)$ (Fig.~\ref{fig:fig 2}b):
\begin{equation}
I_\textrm{C}= I_\textrm{AB}(r)\left[1+\frac{ I_\textrm{T}(r,\theta)-(I_\textrm{A}(r,\theta)+I_\textrm{B}(r,\theta))}{2\sqrt{I_\textrm{A}(r,\theta)I_\textrm{B}(r,\theta)}}\right],
\label{eq:correction}
\end{equation}
where $I_\textrm{AB}(r)=\left(\overline{I_\textrm{A}}(r)+\overline{I_\textrm{B}}(r)\right)/2,$ with $\overline{I_\textrm{A}}(r)$ and $\overline{I_\textrm{B}}(r)$ the average radial profile of $I_\textrm{A}$ and $I_\textrm{B},$ respectively.  For further details see Supplementary Note 2.  

The form of Eq.~\ref{eq:intIncoh} means that the corrected interferogram is readily Fourier analysed.  Obtaining the azimuthal profile by integrating $I(r,\theta)$ (Fig.~\ref{fig:fig 2}c), then performing a one-dimensional Fourier transform, results in a Fourier spectrum where terms with $2\ell$ spatial frequency give $P^{\vert \ell\vert}=\sum_{p} P^{\vert \ell\vert}_p$, for $\vert\ell\vert>0$.  Our current method is limited to measuring $|\ell|$, but $\ell$ can be inferred from OAM conservation \cite{Akulshin2016}.  

Unlike the $\ell$-decomposition, which arises from the rotational symmetry of the beam, the $p$-decomposition is not uniquely defined, but depends on the assumed waist (Gaussian $e^{-2}$ radius, Eq.~\ref{eq:LG field}).  We choose a beam waist such that one `target' mode dominates.  We find the $p$-decomposition for each $\ell$ by separating the corrected interferogram into $\ell$-components using two-dimensional Fourier filtering.  The radial profile of each component is then fitted with an incoherent superposition of $p$-modes (with relevant $\ell$).  

Combining the information from the $\ell$ and $p$ decompositions, the relative power in all modes with $\vert\ell\vert>0$ is found.  To determine the $\ell=0$ $p$-mode populations a final fit is performed to the radial profile of the corrected interferogram (Fig.~\ref{fig:fig 2}d).  The model for the fit is the total $\vert\ell\vert>0$ radial profile, multiplied by a single scale factor, plus an incoherent $\ell=0$ mode sum.  To verify our analysis we use the mode decomposition results to calculate a reconstructed angular and radial profile (Fig.\ \ref{fig:fig 2} c and d), finding good quantitative agreement between theory and experiment, and high purity in the desired specific LG$_p^{\ell}$ mode.

\begin{figure}[!t]
\centering
\includegraphics[width=\linewidth]{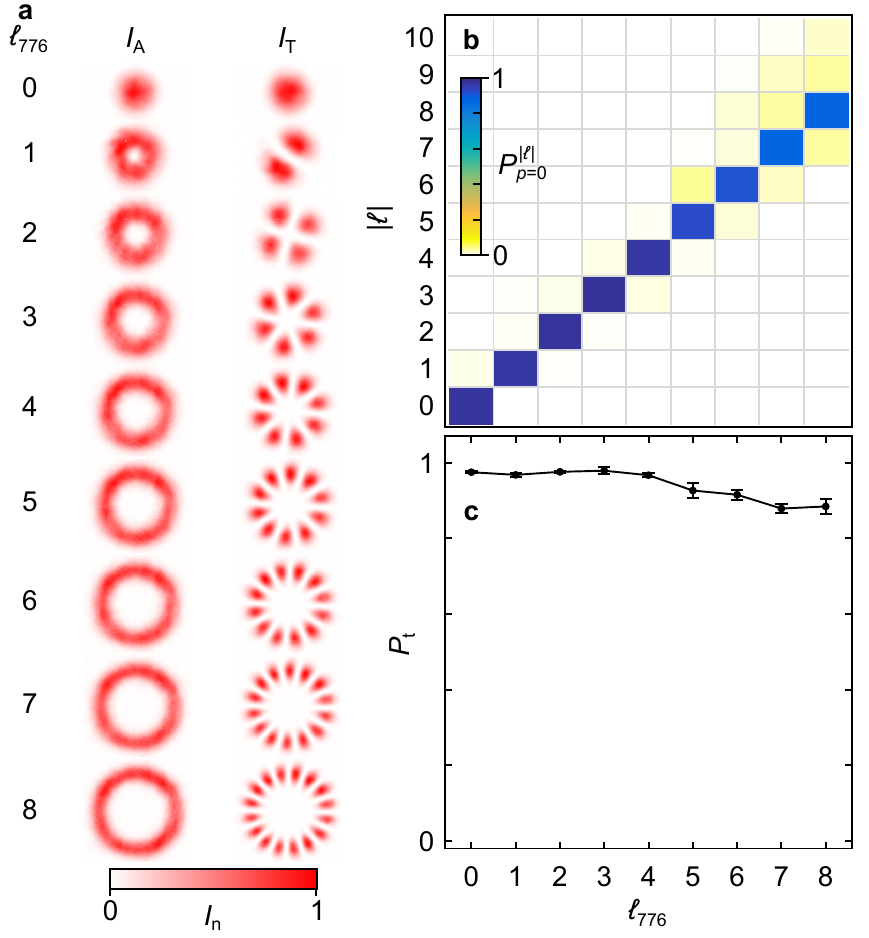}
\caption{Mode decomposition of the 776$\,$nm pump.  (a) Experimental intensity profile in interferometer arm A, $I_\textrm{A}$, and uncorrected interferogram, $I_\textrm{T}$.  (b) Decomposition into $p=0$ modes and (c) relative power $P_\textrm{t}$ in the target mode LG$^{\ell_{776}}_{0}$. Here, and in all subsequent figures, error bars indicate the standard error of the mean based on five measurements.}
\label{fig:fig 3}
\end{figure}

For comparison with the generated $420\,$nm light later, Fig.\ \ref{fig:fig 3}a shows the beam profiles, $I_\textrm{A}$, and uncorrected interferograms, $I_\textrm{T}$, for LG$_{0}^{\ell_{776}}$ pump modes ($\ell_{776}=0 \to 8$) with the resulting mode decompositions in Fig.\ \ref{fig:fig 3}b. Only ${p=0}$ modes are shown, as they constitute ${>96\%}$ of the power in each case, however the full analysis up to $p=3$ is in Supplementary Fig.\ 1 a-d.  The corresponding relative power $P_t$ in the target mode LG$^{\ell_{776}}_{p=0}$ is shown in Fig.\ \ref{fig:fig 3}c.  The measured mode purity is high, $P_t>0.97$ for $\ell_{776}<5$ with a slight decrease as $\ell_{776}$ increases. The 780$\,$nm pump beam has similar mode purities -- see Supplementary Fig.\ 1 e-h.
\\

\begin{figure}[!t]
\centering
\includegraphics[width=\linewidth]{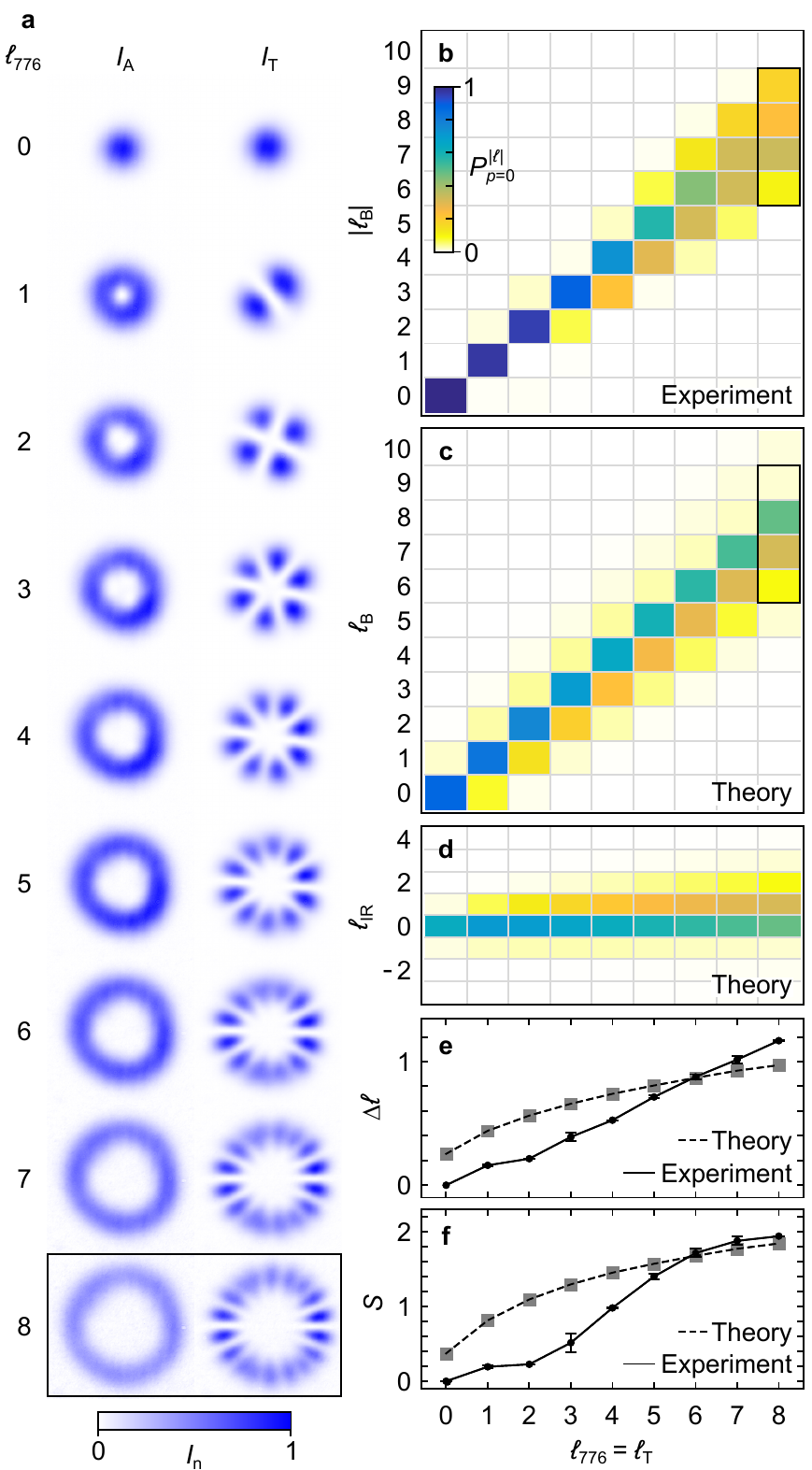}
\caption{Mode decomposition of the 420$\,$nm emission when all pump OAM is carried by the 776$\,$nm field.  (a) Measured intensity profile $(I_\textrm{A})$ and uncorrected interferogram $(I_\textrm{T})$ for each pump mode.  (b) Measured and (c) predicted 420$\,$nm $p=0$ mode decomposition.  (d) Predicted 5.2$\,\mu$m $p=0$ mode decomposition. Experimental (circles) and theoretical (squares) spiral bandwidth $\Delta\ell$ (e) and entanglement entropy $S$ (f) as a function of $\ell_\textrm{T}$ -- curves indicate trends.} 
\label{fig:fig 4}
\end{figure}

\noindent \textbf{Spiral bandwidth broadening.} 
We now consider the FWM experimental results. In the first experiment only the 776$\,$nm pump beam carries OAM, with the 780$\,$nm beam in the Gaussian LG$^0_0$ mode. We measure the 420$\,$nm mode decomposition $P_{p_\textrm{B}}^{\ell_\textrm{B}}$ up to $\ell_\textrm{B}=8$.  The 780$\,$nm beam focus occurs $9.6\,$mm before the focus of the 776$\,$nm beam to improve the spatial overlap of the fields (Methods).  The intensity profiles, $I_\textrm{A}$, and uncorrected interferograms, $I_\textrm{T}$, of the generated 420$\,$nm beams for each pump mode are shown in Fig.\ \ref{fig:fig 4}a. 

Mode decomposition was carried out for each interferogram, considering LG$_{0\leq p\leq 3}^{0\leq\ell\leq 10}$ modes.  We find that $<10\%$ of the light is generated in $p>0$ modes ($<2\%$ theoretically), and present only $p=0$ results here. The experimental and theoretical mode decompositions are in Fig.\ \ref{fig:fig 4} b and c, respectively (see Supplementary Fig.\ 2 a-f for $p>0$ results). 

For low $\ell$ pump modes, $\ell_{776}\leq 3$, the blue light is generated in a single mode, indicated by the high visibility fringes in the interferogram.  In this regime, the results are consistent with nearly all of the pump OAM being transferred to the 420$\,$nm emission.  The 776$\,$nm OAM state is efficiently frequency converted to 420$\,$nm, with measured mode purity $P_t$ inferred in $|$LG$_0^{\ell_\textrm{T}}\rangle_\textrm{B}|$LG$_0^0\rangle_\textrm{IR}$ higher than the theoretical prediction (see Supplementary Fig.\ 2e).  The FWM gain may be sufficient to produce a `lasing' effect, amplifying the optimal mode \cite{Offer2016}.  

The asymmetry in the theoretical generated field OAM mode distributions is due to their large wavelength difference, i.e.\ 420$\,$nm and 5.2$\,\mu$m (Fig.\ \ref{fig:fig 4} c and d), and therefore very different waists. For efficient wave mixing the Rayleigh range of the four fields must be matched --  known for Gaussian beams as the Boyd criterion \cite{Kleinman1968}. We make the assumption that this holds for LG$_p^{\ell}$ modes, and thus the 5.23$\,\mu$m and 420$\,$nm field waists satisfy: 
\begin{equation}
w_\textrm{IR}\simeq\sqrt{\frac{523}{78}}w_{780},\;\; w_\textrm{B}\simeq\sqrt{\frac{42}{78}}w_{780}, \;\; w_{776} \simeq w_{780},
\label{eq:waists}
\end{equation}
approximately 2.6 and 0.73 times the pump waist. Thus for low $\ell_{776}$ there is much better overlap between the fields (which have intensity maxima at radius $\sqrt{\ell/2} w$ \cite{ferris}) if the 5.2$\,\mu$m field is in the Gaussian $\ell=0$ mode \cite{Walker2012} (Fig.\ \ref{fig:fig 4}d -- see Supplementary Fig.\ 2 a-f for $p>0$ mode powers).

As $\ell_{776}$ increases, the 420$\,$nm light spreads over an increasing number of modes -- apparent in the decreasing fringe visibility at the top and bottom of the interferograms (12 and 6 O'clock in Fig.\ \ref{fig:fig 4}a).  The interferogram positions of high visibility (3 and 9 O'clock in Fig.\ \ref{fig:fig 4}a) are determined by the interferometer geometry and in particular the Dove prism orientation (Fig.~\ref{fig:fig 1}) which is used to create the beam's mirror image. Our Dove prism `mirror' is aligned about the horizontal axis where nearby parts of the beam interfere, therefore each separate mode's interferogram is necessarily in phase along the horizontal axis.

When the 420$\,$nm light is observed in a range of modes, due to OAM conservation, the 5.2$\,\mu$m field can no longer be solely generated in the LG$_0^0$ mode.  The radius of an LG mode increases with $\sqrt{\ell}$  so as $\ell_{776}$ increases the overlap with the higher order 5.2$\,\mu$m modes improves.  The two-photon 5.2$\,\mu$m and 420$\,$nm field is then generated as a coherent superposition of different combinations of OAM-conserving modes.  The $\ell_{776}\geq 4$ results, with power spread over more than one 420$\,$nm mode, strongly indicate that generated 5.2$\,\mu$m and 420$\,$nm photon pairs are OAM-entangled, although further experiments are required to verify this.
The number of modes involved in the entangled state, the spiral bandwidth, depends on the width of the 420$\,$nm $\ell$-decomposition, $\Delta \ell$ (Eq.\ \ref{spiralbandwidth}).  We observe a strongly mode-dependent spiral bandwidth (Fig.\ \ref{fig:fig 4}e) and entanglement entropy (Eq.~\ref{entanglemententropy}, Fig.\ \ref{fig:fig 4}f), that increase with pump OAM.

Finally, we note the effect of Gouy phase matching on the mode decomposition. The Gouy phase, $\Phi_\textrm{G}$ in Eq.~\ref{eq:LG field}, describes a modification of the phase of a focused beam compared to that of a collimated beam. In order for phase matching to be maintained between the pump and generated beams throughout the cell, they must have identical Gouy phases - a requirement we call Gouy-phase matching.  This means that the mode order $2p+\vert\ell\vert$ must be conserved, i.e.\ for $p=0$ pump modes, the only Gouy-phase matched generated modes satisfy:
\begin{equation}
\vert\ell_{780}\vert+\vert\ell_{776}\vert=\vert\ell_\textrm{IR}\vert+2 p_\textrm{IR}+\vert\ell_\textrm{B}\vert+2 p_\textrm{B}.
\label{eq:gpm}
\end{equation}

\begin{figure}[!t]
\centering
\includegraphics[width=\linewidth]{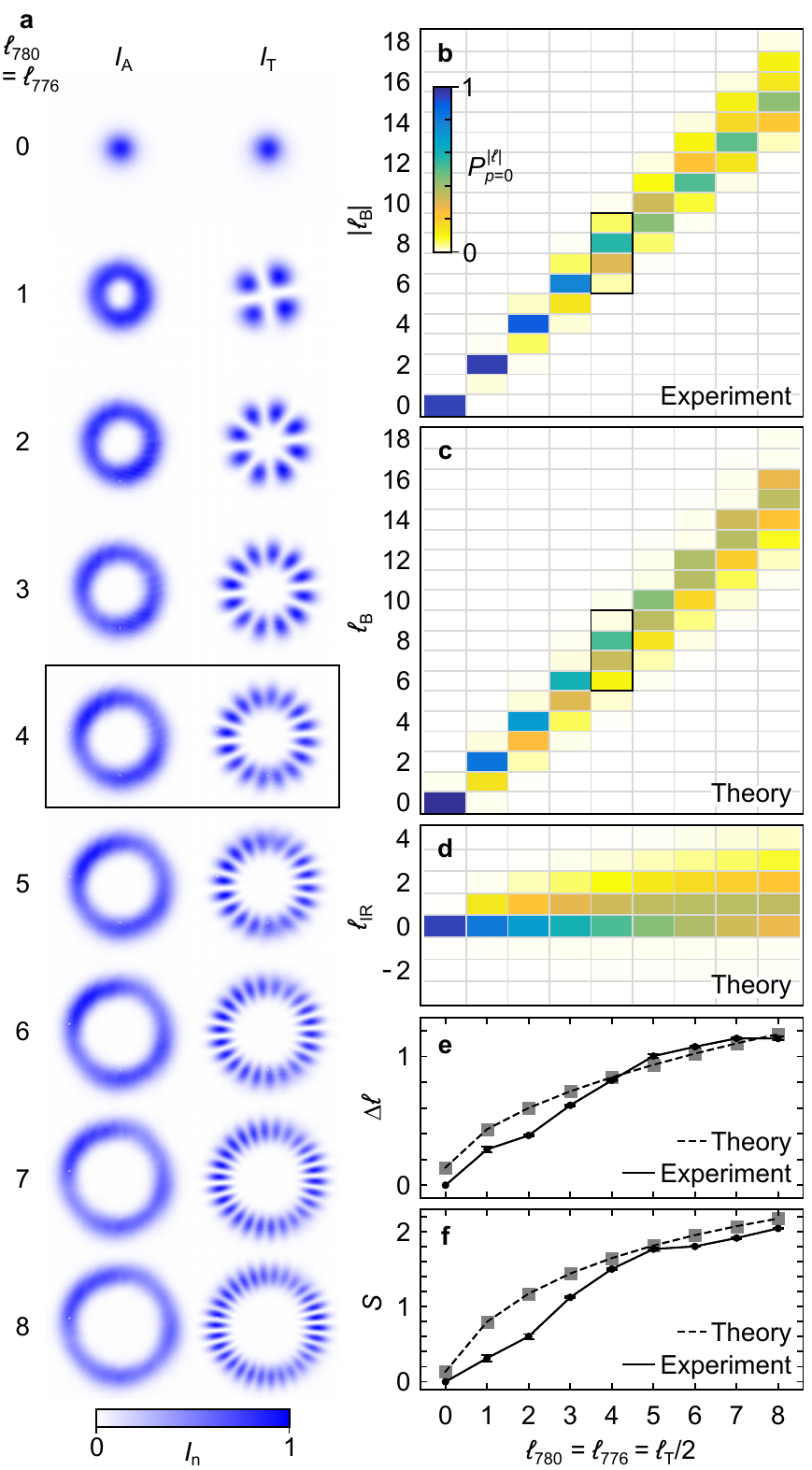}
\caption{The 420$\,$nm mode decomposition when OAM is shared between the pump fields.  (a) Experimental uncorrected interferogram, $I_\textrm{T}$, and intensity profile, $I_\textrm{A}$, for each mode.  (b) Measured and (c) predicted 420$\,$nm $p=0$ mode decomposition.  (d) Predicted 5.2$\,\mu$m $p=0$ mode decomposition. Experimental (circles) and theoretical (squares) spiral bandwidth $\Delta\ell$ (e) and entanglement entropy $S$ (f) as a function of $\ell_\textrm{T}/2$ -- curves indicate trends.} 
\label{fig:fig 5}
\end{figure}

Both the theoretical and experimental results show a 420$\,$nm $\ell$-decomposition asymmetry; there is relatively more power in modes with $\ell$ lower than that of the mode LG$^{\ell_{776}}_{p=0}$.  To conserve OAM, if the 420$\,$nm light has $\ell>\ell_{776}=\ell_\textrm{T}$, then the 5.2$\,\mu$m mode must have negative $\ell$.  Considering Eq.\ \ref{eq:gpm} this situation cannot be well Gouy phase matched (since $p\geq0$), and therefore modes with $\ell>\ell_{776}$ are less likely. With axially offset 780$\,$nm and 776$\,$nm foci, the 780$\,$nm Gouy phase is essentially constant through the FWM region.  This is included in our theoretical model (Methods), and reduces the effect of Gouy phase matching compared to when both beams are confocal in the next section of the paper.
\\

\noindent\textbf{Shared pump OAM.}
Na\"{i}vely one might think it doesn't matter whether the OAM is provided by only one or both of the pump beams.  However, we observe a reduction of spiral bandwidth for a given target mode if both pump beams carry OAM.  In our second FWM experiment the two pump beams are shaped into the same LG mode, $\ell_{780}=\ell_{776}=\ell_\textrm{T}/2,$ $p_{780}=p_{776}=0$, overlapped and  focused into the centre of the rubidium cell.  Fig.\ \ref{fig:fig 5}a shows the  intensity profile, $I_\textrm{A}$, and uncorrected interferogram, $I_\textrm{T}$, of the generated 420$\,$nm beam for each pump $\ell$, whilst Fig.~\ref{fig:fig 5} b-d show the experimental 420$\,$nm and theoretical 420$\,$nm and $5.2\,\mu$m $p=0$ mode decompositions, respectively (see Supplementary Fig.\ 2 g-l for $p>0$).  

Like Fig.~\ref{fig:fig 4}, for low $\ell$ pump beams,  the 420$\,$nm mode decomposition is consistent with all OAM transferred to the 420$\,$nm light, and mode purity $P_\textrm{t}$ inferred in 
$|$LG$_0^{\ell_\textrm{T}}\rangle_\textrm{B}|$LG$_0^0\rangle_\textrm{IR}$ again higher than predicted by theory (see Supplementary Figure 2k). Here, the blue light OAM is the sum of the two pump beam OAMs, and we demonstrate efficient OAM addition. As the pump $\ell$ increases we see the 420$\,$nm power spread over more modes and the available spiral bandwidth (Fig.\ \ref{fig:fig 5}e) and entanglement entropy increase (Fig.\ \ref{fig:fig 5}f), both in the experiment and  theory.  

A comparison of providing OAM from one or both pump beams can be seen in Figs.~\ref{fig:fig 4} and \ref{fig:fig 5}, where the identical  blue mode LG$^8_0$ is highlighted by boxes in both cases. We observe a larger spread of 420$\,$nm modes, indicating a larger spiral bandwidth, when all of the OAM is provided by the 776$\,$nm field (Fig.\ \ref{fig:fig 4}) than when the pump OAM is shared equally between the two pump fields (Fig.\  \ref{fig:fig 5}).  This is theoretically expected due to increased mode overlap for $\ell_{780}=\ell_{776}$, however experimentally we observe an even larger discrepancy -- see Supplementary Note 5 and Supplementary Fig.\ 5 for more details.  
Comparing Figs.~\ref{fig:fig 4} and \ref{fig:fig 5} also shows the relative importance of Gouy phase matching: in Fig.\ \ref{fig:fig 4} and \ref{fig:fig 5} the pump beams have foci that are  displaced along the beam axis, or co-located, respectively.  The increased pump beam propagation symmetry in the latter case leads to stronger Gouy phase matching.  As a result, both experimentally and theoretically, there is stronger asymmetry in the 420$\,$nm $\ell$-distribution about the target mode in Fig.\ \ref{fig:fig 5} than in Fig.\ \ref{fig:fig 4}.  This is also apparent in the predicted 5.2$\,\mu$m mode decomposition, where essentially no light is predicted with $\ell<0$ in Fig.\ \ref{fig:fig 5}, but some is predicted in Fig.\ \ref{fig:fig 4}.\\

\begin{figure}[!t]
\centering
\includegraphics[width=\linewidth]{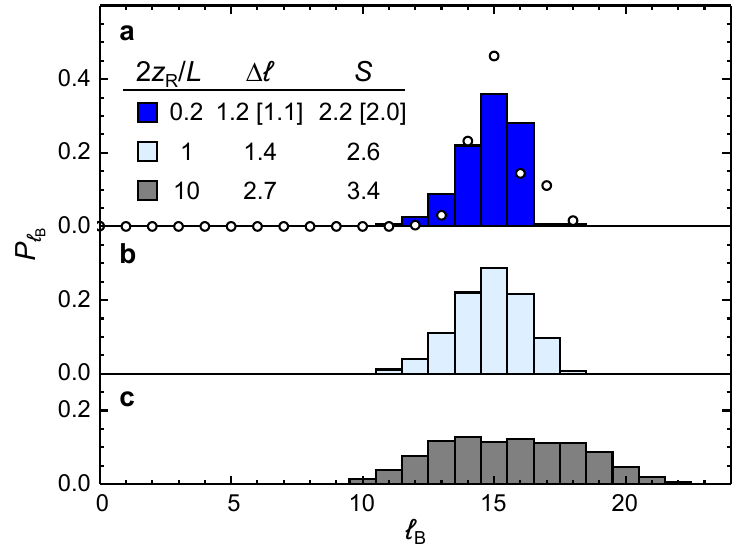}
\caption{Theoretically predicted blue $\ell_\textrm{B}$ distribution, spiral bandwidth  $\Delta \ell(\ell_\textrm{T})$ (Eq.~\ref{spiralbandwidth}) and entanglement entropy $S(\ell_\textrm{T})$ (Eq.~\ref{entanglemententropy}) for a total OAM of $\ell_\textrm{T}=16.$ Double Rayleigh:cell length parameters $2 z_\textrm{R}/L$ of 0.2, 1 and 10 are considered in the  histograms a, b and c respectively. Note that the $2 z_\textrm{R}/L=0.2$ results (circles: experimental values), and theoretical [experimental] $\Delta \ell(\ell_\textrm{T})$ and $S(\ell_\textrm{T})$ values, refer to the distributions in the rightmost columns of Fig.~\ref{fig:fig 5}c [b], which have similar widths and trends.}
\label{fig:fig 6}
\end{figure}

\noindent\textbf{Cell vs.\ Rayleigh length effects.} In our experiments we work in a regime, $2 z_\textrm{R}/L=0.2,$ where the cell length $L$ is significantly larger than the focal region of the pump beams $\sim 2 z_\textrm{R}.$ It is instructive to consider how the spiral bandwidth might change in situations with high-power pumps under a weaker focus, as in squeezing experiments\cite{irina} and related blue-light FWM experiments\cite{Pruvost2018}. In Fig.~\ref{fig:fig 6} we show a comparison between the theoretical and experimental spiral bandwidth and entanglement entropy results for our system.

We also theoretically consider conditions under which the double Rayleigh range vs.\ cell length parameter is around unity\cite{Pruvost2018,irina}, or ten. Moreover, thin cells can be used for Faraday filters\cite{ifan}, with ultrathin cells for collective Lamb shift\cite{collective} and Rydberg\cite{rydberg} experiments, and as we see in Fig.~\ref{fig:fig 6} short cells also offer 
prospects for increasing the spiral bandwidth of entangled light in FWM systems. We stress that these results solely rely on the LG mode overlap integrals, and could be quantitatively altered by any propagation effects like gain/absorption and Kerr lensing on any of the involved FWM transitions, however we expect that the trend of increasing spiral bandwidth and entanglement entropy for increasing $2 z_\textrm{R}/L$ should hold.
\\

\noindent {\large {\bf \textsf{Discussion}}}
\\
We have developed a method of precisely determining the  decomposition of incoherent beam modes and used this to analyse the transverse light created via FWM in Rb vapour.  Due to the large wavelength difference of the two generated fields, the OAM distribution between them is pump-mode dependent.  A simple theoretical model \cite{Walker2012,Lanning2017}, which we adapted to include a full $\ell$ and $p$ description of all four fields, describes our system well, but could be improved by including propagation effects and a more accurate generated field waist prediction.  

For small pump OAM $(\ell\leq 3)$, our results are consistent with all OAM being transferred to the 420$\,$nm light.  In this regime the FWM process can be faithfully used to both frequency convert and add OAM states.  As the pump OAM increases the 420$\,$nm light is generated in an incoherent superposition of an increasing number of modes.  This indicates that the pump OAM is shared between the 5.2$\,\mu$m and 420$\,$nm light, which is likely to lead to highly efficient generation of 
OAM-entangled photon pairs. 
The inferred spiral bandwidth and entanglement entropy in our system is relatively modest, however we find that they increase with increasing pump OAM and predict further increases for shorter cell lengths.  

In addition, the OAM distribution between the generated fields depends on how the pump OAM is supplied.  We see a significant decrease in spiral bandwidth when the pump OAM is shared equally between the pump beams rather than being supplied by only one pump beam.  Our results also show the importance of Gouy phase matching, which results here in an asymmetric $\ell_\textrm{B}$-distribution about the $|$LG$_0^{\ell_\textrm{T}}\rangle_\textrm{B}$ mode.

As well as demonstrating a means of manipulating and generating OAM states, these results are relevant for schemes involving inscription and storage of phase information in atomic gases.  Although we currently work with an atomic vapour, the techniques presented here are well-suited for use with cold atoms \cite{Nicolas2014,Ding2015}, which will enable long lifetime high-dimensional quantum memories with wavelength versatility.  Finally, we note the 420$\,$nm output could be constrained to a single transverse mode, even for high input OAM, if a cavity\cite{Offer2016,Brekke2016} were to be added to the system.  The output is then determined by a combination of the coherent FWM pumping mechanism and the cavity modes. \\

\noindent {\large {\bf \textsf{Methods}}}
\\
{\footnotesize
\noindent \textbf{Energy levels and wavelengths.}
Two near-infrared pump fields (780$\,$nm and 776$\,$nm) generate an infrared field (5.2$\,\mu$m) and a blue field (420$\,$nm).  With Gaussian pump beams, this resonantly enhanced process can be carried out very efficiently \cite{Zibrov2002,Meijer2006,Akulshin2009,Vernier2010}, generating mW levels of 420$\,$nm light with low power pump beams \cite{Vernier2010}.  Although we use single pass FWM here, the process can also be enhanced through the use of a cavity resonant with either the generated \cite{Offer2016} or pump light \cite{Brekke2016}. 

\noindent \textbf{Laser locks.}
In order to minimise single-photon absorption \cite{Vernier2010}, the 780$\,$nm laser \cite{Arnold1998} is locked roughly half way between the hyperfine states of the $^{85}$Rb 5S$_{1/2}$ ground state (+1.6$\,$GHz from the 5S$_{1/2}$ $F=3$ to 5P$_{3/2}$ $F'=4$ transition). The output power in each of the fiber-coupled $780\,$nm and $776\,$nm pump beams  (Fig.~\ref{fig:fig 1}) was around $30\,$mW. Typically about $1\,$mW, with a relatively weak $\ell$ dependence, was converted into almost pure LG pump modes after the SLM. The 776$\,$nm laser is locked two-photon resonant with the $^{85}$Rb 5S$_{1/2}$ $F=3$ to 5D$_{5/2}$ $F'=5$ transition. The auto-correlation linewidths of the 780, 776$\,$nm  lasers are 0.6, 0.2$\,$MHz, respectively, on a $0.1\,$ms timescale \cite{Offer2016} and we expect similar values for $10\,$ms timescales \cite{Arnold1998}. The two generated fields are quasi-resonant with the downward cascade via the 6P$_{3/2}$ state \cite{Akulshin2012,Walker2012,Offer2016}. Note that the blue beam powers of 400-10$\,$nW, typically decreasing with $\ell,$ indicate that the `spacing' between blue photons is up to $1.5\,$cm -- the blue light power is only two orders of magnitude away from the single-photon regime.    

\noindent \textbf{The optical setup.}
The two pump fields are shaped independently (Fig.~\ref{fig:fig 1}) into a range of LG modes by displaying the corresponding holograms each on one half of a spatial light modulator (SLM, Hamamatsu LCOS-SLM X13138). The required phase holograms are calculated following method C outlined in Refs.\ \citen{Clark2016,Radwell2017}, and first proposed in Ref.~\citen{Davis1999}.  The shaped pump beams are combined on a non-polarising beam splitter (NPBS), circularly polarised, and focused to a $w_0=25\,\mu$m waist at the centre of a 120$^{\circ}$C Rb cell ($P_\textrm{Rb}\sim 9\times 10^{-4}\,$mbar). The Rayleigh range $z_\textrm{R}=2.5\,$mm is much less than the $25\,$mm cell length. FWM generates light at $420\,$nm and also $5.2\,\mu$m.  

\noindent \textbf{Mode decomposition.}
The transverse mode decomposition of the pump and generated fields are determined from interferograms  obtained by overlapping a beam with its mirror image.  A single field is chosen with a spectral filter (for the 420$\,$nm field) or by blocking one of the pump beams, and the light is then linearly polarised.  The light passes through a Mach-Zehnder interferometer, with a Dove prism in one of its arms to generate the light mode's mirror image. The resulting interference pattern after the final NPBS contains azimuthal lobes, when carefully aligned.  Interferogram analysis gives the $\ell$ and $p$-mode decomposition of each of the fields.  See Supplementary Note 2 for further experimental details.

\noindent\textbf{Overlap integral.} 
The generated field $\ell$ and $p$ mode decomposition for a particular pair of pump modes can be predicted by considering the FWM field overlap within the Rb cell \cite{Walker2012,Lanning2017}. The probability amplitude to generate a particular pair of 5.2$\,\mu$m and 420$\,$nm modes, when pumping with  LG$_{p_{780}}^{\ell_{780}}$ and LG$_{p_{776}}^{\ell_{776}}$, can be found by evaluating the integral: 
\begin{equation}
c_{p_\textrm{B}, p_\textrm{IR}}^{\ell_\textrm{B}, \ell_\textrm{IR}}= \int\limits_{-L/2}^{L/2}\int\limits_0^R\int\limits_0^{2\pi}r \textrm{LG}_{p_{780}}^{\ell_{780}*}\textrm{LG}_{p_{776}}^{\ell_{776}*}\textrm{LG}_{p_\textrm{B}}^{\ell_\textrm{B}}\textrm{LG}_{p_\textrm{IR}}^{\ell_\textrm{IR}}d\theta drdz ,
\label{eq:integral}
\end{equation}
where the mode waists are given by Eq.\ \ref{eq:waists} and $L$ is the cell length \cite{Walker2012,Lanning2017}.  
The azimuthal integral ensures OAM $(\ell)$, is conserved
\begin{equation}
\ell_{780}+\ell_{776}=\ell_\textrm{IR}+\ell_\textrm{B},
\label{eq:lcons}
\end{equation}
whilst the radial integral considers the field spatial overlap.  The $z$ integral specifies the beam propagation through the cell.

\noindent \textbf{Theory of the generated fields.} We assume that the 5.2$\,\mu$m and 420$\,$nm light is generated as a coherent superposition of two-photon states, each weighted by coefficients $c_{p_\textrm{B}, p_\textrm{IR}}^{\ell_\textrm{B}, \ell_\textrm{IR}}$
\begin{equation}
E^{2Photon}=\sum\limits_{\ell_\textrm{IR},p_\textrm{IR}}  \sum\limits_{\ell_\textrm{B},p_\textrm{B}} c_{p_\textrm{B}, p_\textrm{IR}}^{\ell_\textrm{B}, \ell_\textrm{IR}}\textrm{LG}_{p_\textrm{B}}^{\ell_\textrm{B}}\textrm{LG}_{p_\textrm{IR}}^{\ell_\textrm{IR}} .
\label{eq:2PhotonState}
\end{equation}
Although the two-photon field is coherent, the 420$\,$nm field alone is an incoherent mixture of modes.  We obtain the probability of observing the 420$\,$nm light in a specific mode by summing over all 5.2$\,\mu$m modes  
$P_{p_\textrm{B}}^{\ell_\textrm{B}}=\sum\limits_{\ell_\textrm{IR},p_\textrm{IR}} 
\left\vert c_{p_\textrm{B}, p_\textrm{IR}}^{\ell_\textrm{B}, \ell_\textrm{IR}} \right\vert^2.$ The model neglects the propagation effects of absorption and Kerr lensing -- a reasonable assumption as the 780$\,$nm laser is detuned +1.6$\,$GHz and -1.4$\,$GHz from the $F=3$ and $F=2$ ground states, respectively \cite{Vernier2010}.  

In the experiment, we compare FWM under two conditions:  all pump OAM carried only by the 776$\,$nm field (with the 780$\,$nm field in the LG$_{0}^{0}$ mode); and with the OAM shared evenly between the pump beams.  In the first case we axially offset the 780$\,$nm field focus to $z_{\mathrm{off}}=9.6\,$mm (3.8$z_\textrm{R}$) before that of the 776$\,$nm field, to improve the spatial overlap of the higher order 776$\,$nm modes with the Gaussian 780$\,$nm beam.  The axial offset is included in the model by performing the transformation $z\rightarrow z+z_{\mathrm{off}}$ on the 780$\,$nm mode before evaluating Eq.~\ref{eq:integral}.  This causes a mismatch in the pump beam phase front curvature, but we assume overall phase-front matching of the two pump fields with the generated two-photon field.  The offset model is in agreement with our experimental observations; we observe a change in the 420$\,$nm field collimation when the 780$\,$nm focus shifts.

\noindent\textbf{Data availability.} 
The datasets used in this work are available via the Pure repository (DOI: 10.15129/96db0ebb-aace-494f-8e61-d4a064fcadbb)\onlinecite{Offer2018}.\\

\noindent {\large {\bf \textsf{References}}}\vspace{-1cm}

\noindent \\
{\large \textbf{\textsf{Acknowledgments}}}\\
We are grateful for funding from the Leverhulme Trust (RPG-2013-386) and EPSRC (EP/M506643/1). We thank J.\ W.\ C.\ Conway for initial work with the SLM.  

\noindent \\
{\large \textbf{\textsf{Author contributions}}}\\
S.F.-A.\ and A.S.A.\ devised the scheme and led the research, with input from E.R. R.F.O.\ performed the experiments and wrote the draft manuscript. R.F.O.\ implemented the theoretical model, extending the initial model of D.S. All authors discussed the results and implications, and commented
on the manuscript.

\noindent \\
{\large \textbf{\textsf{Additional information}}}\\
\noindent \textbf{Supplementary information} 
accompanies this article.

\noindent \textbf{Competing interests:} The authors declare no competing interests.


\begin{thebibliography}{10}
{\footnotesize

\bibitem{Franke-Arnold2008}
Franke-Arnold, S., Allen, L.\ \& Padgett, M.\ Advances in optical angular momentum.\ 
\dois{10.1002/lpor.200810007}{\textit{Laser Photon.\ Rev.} \textbf{2}, 299--313 (2008).} 

\bibitem{barnett}
Barnett, S.~M.\ Relativistic electron vortices.\ 
\dois{10.1103/PhysRevLett.118.114802}{\textit{Phys. Rev. Lett.} \textbf{118}, 114802 (2017)}.

\bibitem{clark}
Clark, C.~W., Barankov, R., Huber, M.~G., Arif, M., Cory, D.~G.\ \&  Pushin, D.~A.\ Controlling neutron orbital angular momentum.\ 
\dois{10.1038/nature15265}{\textit{Nature} \textbf{525}, 504--506 (2015)}. 

\bibitem{Franke-arnold2017}
Franke-Arnold, S.\ \& Radwell, N.\ Light served with a twist.\ 
 \dois{10.1364/OPN.28.6.000028}{\textit{Opt.\ Phot.\ News} \textbf{28}, 28--35 (2017)}.
\bibitem{Wang2012}

Wang, J.\ \textit{et al}.\ 
Terabit free-space data transmission employing orbital angular momentum multiplexing.\ 
  \dois{10.1038/nphoton.2012.138}{\textit{Nature Phot}.\ \textbf{6}, 488--496 (2012)}.

\bibitem{Langford2004}
Langford, N.~K.\ \textit{et al}.\ 
  Measuring entangled qutrits and their use for quantum bit commitment.\ 
  \dois{10.1103/PhysRevLett.93.053601}{\textit{Phys.\ Rev.\ Lett}.\   \textbf{93}, 053601 (2004)}.

\bibitem{Groblacher2006}
Gr{\"{o}}blacher, S.\ \textit{et al}.\ 
 Experimental quantum cryptography with qutrits.  
  \dois{10.1088/1367-2630/8/5/075}{\textit{New J.\ Phys}.\ \textbf{8}, 75 (2006)}.

\bibitem{Mafu2013}
Mafu, M.\ \textit{et al}.\ 
  Higher-dimensional orbital-angular-momentum-based quantum key distribution with mutually unbiased bases.\
  \dois{10.1103/PhysRevA.88.032305}{\textit{Phys.\ Rev.\ A} \textbf{88}, 032305 (2013)}.

\bibitem{Dholakia1996}
Dholakia, K., Simpson, N.~B., Padgett, M.~J.\ \& Allen, L.\
 Second-harmonic generation and the orbital angular momentum of light.\  \dois{10.1103/PhysRevA.54.R3742}{\textit{Phys.\ Rev.\ A} \textbf{54}, R3742 (1996)}.

\bibitem{Mair2001}
Mair, A., Vaziri, A., Weihs, G.\ \& Zeilinger, A.\ 
Entanglement of the orbital angular momentum states of photons. 
  \dois{10.1038/35085529}{\textit{Nature} \textbf{412}, 313--316 (2001)}.

\bibitem{Steinlechner2015}
Steinlechner, F., Hermosa, N., Pruneri, V.\ \& Torres, J.~P.\ 
Frequency conversion of structured light.\  
  \dois{10.1038/srep21390}{\textit{Sci.\ Rep}.\ \textbf{6}, 21390 (2015)}.

\bibitem{Ding2012}
Ding, D.~S.\ \textit{et al}.\ 
  Experimental up-conversion of images.\   
  \dois{10.1103/PhysRevA.86.033803}{\textit{Phys.\ Rev.\ A} \textbf{86}, 033803 (2012)}.

\bibitem{Zhou2016}
Zhou, Z.~Y.\ \textit{et al}.\ 
   Orbital angular momentum-entanglement frequency transducer.   \dois{10.1103/PhysRevLett.117.103601}{\textit{Phys.\ Rev.\ Lett}.\ \textbf{117}, 103601 (2016)}.

\bibitem{Li2015}
Li, Y., Zhou, Z.-Y., Ding, D.-S.\ \&  Shi, B.-S.\
 Sum frequency generation with two orbital angular momentum carrying laser beams.\ 
  \dois{10.1364/JOSAB.32.000407}{\textit{J.\ Opt.\ Soc.\ Am.\ B} \textbf{32}, 407--411 (2015)}.

\bibitem{Walker2012}
Walker, G., Arnold, A.~S.\ \& Franke-Arnold, S.\ 
 Trans-spectral orbital angular momentum transfer via four-wave mixing in Rb vapor.\
  \dois{10.1103/PhysRevLett.108.243601}{\textit{Phys.\ Rev.\ Lett}.\ \textbf{108}, 243601 (2012)}.

\bibitem{Akulshin2016}
Akulshin, A.~M., Novikova, I., Mikhailov, E.~E., Suslov, S.~A.\ \&  McLean, R.~J.\
  Arithmetic with optical topological charges in stepwise-excited Rb vapor.\  
  \dois{10.1364/OL.41.001146}{\textit{Opt.\ Lett}.\ \textbf{41}, 1146--1149 (2016)}.

\bibitem{Pruvost2018}
  Chopinaud, A., Jacquey, M., Viaris de Lesegno, B.\ \& Pruvost L.\
	High helicity vortex conversion in a rubidium vapor.\ 
  \dois{10.1103/PhysRevA.97.063806}{\textit{Phys.\ Rev.\ A} \textbf{97}, 063806  (2018)}.

\bibitem{Franke-Arnold2002}
Franke-Arnold, S., Barnett, S.~M., Padgett, M.~J.\ \& Allen, L.\ 
 Two-photon entanglement of orbital angular momentum states.\ 
  \dois{10.1103/PhysRevA.65.033823}{\textit{Phys.\ Rev.\ A} \textbf{65}, 033823 (2002)}.

\bibitem{Boyer2008}
Boyer, V., Marino, A.~M., Pooser, R.~C.\ \& Lett, P.~D.\
 Entangled images from four-wave mixing.\  
  \dois{10.1126/science.1158275}{\textit{Science} \textbf{321}, 544--547 (2008)}.

\bibitem{Marino2009}
Marino, A.~M., Pooser, R.~C., Boyer,  V.\ \& Lett, P.~D.\
 Tunable delay of Einstein-Podolsky-Rosen entanglement.\   \dois{10.1038/nature07751}{\textit{Nature} \textbf{457}, 859--862 (2009)}.
 
\bibitem{Shi2018}
Ding, D.-S., Dong, M.-X., Zhang, W., Shi, S., Yu, Y.-C., Ye, Y.-H., Guo, G.-C. \& Shi, B.-S.\ Experimental demonstration of quantum wrenching orbital angular momentum memory.\ \href{https://arxiv.org/abs/1806.10407}{\textit{arXiv:1806.10407} (2018)}.

\bibitem{Ma2017}
Ma, R., Liu, W., Qin, Z., Jia, X.\ \& Gao, J.\ 
 Generating quantum correlated twin beams by four-wave mixing in hot cesium vapor.\ 
  \dois{10.1103/PhysRevA.96.043843}{\textit{Phys.\ Rev.\ A} \textbf{96}, 043843 (2017)}.

\bibitem{Ihn2017}
Ihn, Y.~S., Park, K.-K., Kim, Y., Chough, Y.-T.\ \& Kim, Y.-H.\ 
 Intensity correlation in frequency upconversion via four-wave mixing in rubidium vapor.\  
 \dois{10.1364/JOSAB.34.002352}{\textit{J.\ Opt.\ Soc.\ Am.\ B} \textbf{34}, 2352--2357 (2017)}.

\bibitem{Whiting2017}
Whiting, D.~J., {\v{S}}ibali{\'{c}}, N., Keaveney, J., Adams, C.~S.\ \&  Hughes, I.~G.\ 
 Single-photon interference due to motion in an atomic collective excitation.\ 
  \dois{10.1103/PhysRevLett.118.253601}{\textit{Phys.\ Rev.\ Lett}.\ \textbf{118}, 253601 (2017)}.

\bibitem{Chaneliere2006}
Chaneli{\`{e}}re, T.\ \textit{et al}.\ 
 Quantum telecommunication based on atomic cascade transitions.\ 
  \dois{10.1103/PhysRevLett.96.093604}{\textit{Phys.\ Rev.\ Lett}.\ \textbf{96}, 093604 (2006)}.

\bibitem{Zibrov2002}
Zibrov, A.~S., Lukin, M.~D., Hollberg, L.\ \& Scully, M.~O.\ 
 Efficient frequency up-conversion in resonant coherent media.\   \dois{10.1103/PhysRevA.65.051801}{\textit{Phys.\ Rev.\ A} \textbf{65}, 051801 (2002)}.

\bibitem{Meijer2006}
Meijer, T., White, J.~D., Smeets, B., Jeppesen, M.\ \& Scholten, R.~E.\ 
 Blue five-level frequency-upconversion system in rubidium.\  
  \dois{10.1364/OL.31.001002}{\textit{Opt.\ Lett}.\ \textbf{31}, 1002--1004 (2006)}.

\bibitem{Akulshin2009}
Akulshin, A.~M.,  McLean, R.~J., Sidorov, A.~I.\ \& Hannaford, P.\ 
 Coherent and collimated blue light generated by four-wave mixing in Rb vapour.\  
 \dois{10.1364/OE.17.022861}{\textit{Opt.\ Express} \textbf{17}, 22861--22870 (2009)}.

\bibitem{Vernier2010}
Vernier, A., Franke-Arnold, S., Riis, E.\ \& Arnold, A.~S.\
 Enhanced frequency up-conversion in Rb vapor.\ 
  \dois{10.1364/OE.18.017020}{\textit{Opt.\ Express} \textbf{18}, 17020--17026 (2010)}.

\bibitem{Becerra2008}
Becerra, F.~E., Willis, R.~T., Rolston, S.~L.\ \& Orozco, L.~A.\ 
 Nondegenerate four-wave mixing in rubidium vapor: The diamond configuration.\ 
  \dois{10.1103/PhysRevA.78.013834}{\textit{Phys.\ Rev.\ A} \textbf{78}, 013834 (2008)}.

\bibitem{Srivathsan2013}
Srivathsan, B.\ \textit{et al}.\ 
 Narrow band source of transform-limited photon pairs via four-wave mixing in a cold atomic ensemble.\ 
 \dois{10.1103/PhysRevLett.111.123602}{\textit{Phys.\ Rev.\ Lett}.\ \textbf{111}, 123602 (2013)}.

\bibitem{Whiting2018}
Whiting, D.~J., Mathew, R.~S., Keaveney, J., Adams, C.~S.\ \& Hughes, I.~G.\ 
 Four-wave mixing in a non-degenerate four-level diamond configuration in the hyperfine Paschen–Back regime.\ 
  \dois{10.1080/09500340.2017.1377308}{\textit{J. Mod. Opt}.\ \textbf{65}, 713--722 (2018)}.

\bibitem{Lanning2017}
Lanning, R.~N.\ \textit{et al}.\ 
 Gaussian-beam-propagation theory for nonlinear optics
  involving an analytical treatment of orbital-angular-momentum transfer.\ 
  \dois{10.1103/PhysRevA.96.013830}{\textit{Phys.\ Rev.\ A} \textbf{96}, 013830 (2017)}.

\bibitem{Allen1992}
Allen, L.~Beijersbergen, M.~W. Spreeuw, R.~J.~C.\ \& Woerdman, J.~P.\ 
  Orbital angular momentum of light and the transformation of Laguerre-Gaussian laser modes.\ 
  \dois{10.1103/PhysRevA.45.8185}{\textit{Phys.\ Rev.\ A} \textbf{45}, 8185--8189 (1992)}.

 \bibitem{Akulshin2014a}
Akulshin, A., Budker, D.\ \& McLean, R.\  
 Directional infrared emission resulting from cascade population inversion and four-wave mixing in Rb vapor.\  
  \dois{10.1364/OL.39.000845}{\textit{Opt.\ Lett}.\ \textbf{39}, 845 (2014)}.

\bibitem{Torres2003}
Torres, J.~P., Alexandrescu, A.\ \& Torner, L.\ 
 Quantum spiral bandwidth of entangled two-photon states.\  
  \dois{10.1103/PhysRevA.68.050301}{\textit{Phys.\ Rev.\ A} \textbf{68}, 050301(R) (2003)}.

 \bibitem{shan}
Leach, J., Bolduc, E., Gauthier, D.~J.\ \& Boyd, R.~W.\ Secure information capacity of photons entangled in many dimensions.\ 
\dois{10.1103/PhysRevA.85.060304}{\textit{Phys.\ Rev.\ A} \textbf{85}, 060304(R) (2012)}.

\bibitem{Flamm2012}
Flamm, D., Naidoo, D., Schulze, C., Forbes, A.\ \& Duparre, M.\  
 Mode analysis with a spatial light modulator as a correlation filter.\  
 \dois{10.1364/OL.37.002478}{\textit{Opt.\ Lett}.\ \textbf{37}, 2478--2480 (2012)}. 

\bibitem{Forbes2016}
Forbes, A., Dudley, A.\ \& McLaren, M.\ 
 Creation and detection of optical modes with spatial light modulators.\ 
  \dois{10.1364/AOP.8.000200}{\textit{Adv.\ Opt.\ Phot}. \textbf{8}, 200--207 (2016)}.

\bibitem{Rui2016}
Rui, G., Gu, B., Cui, Y.\ \& Zhan, Q.\ 
 Detection of orbital angular momentum using a photonic integrated circuit.\  
 \dois{10.1038/srep28262}{\textit{Sci.\ Rep}.\ \textbf{6}, 28262 (2016)}.

\bibitem{Leach2002}
Leach, J., Padgett, M., Barnett, S., Franke-Arnold, S.\ \& Courtial, J.\ 
 Measuring the orbital angular momentum of a single photon.\  
 \dois{10.1103/PhysRevLett.88.257901}{\textit{Phys.\ Rev.\ Lett}.\ \textbf{88}, 257901 (2002)}.

\bibitem{Gonzalez2006}
Gonz{\'{a}}lez, N., Molina-Terriza, G.\ \& Torres, J.~P.\
 How a Dove prism transforms the orbital angular momentum of a light beam.\  
 \dois{10.1364/OE.14.009093}{\textit{Opt.\ Express} \textbf{14}, 9093--9102 (2006)}.

\bibitem{Berkhout2010}
Berkhout, G.~C., Lavery, M.~P., Courtial, J., Beijersbergen, M.~W.\ \&   Padgett, M.~J.\ 
  Efficient sorting of orbital angular momentum states of light.\  
  \dois{10.1103/PhysRevLett.105.153601}{\textit{Phys.\ Rev.\ Lett}.\ \textbf{105}, 153601 (2010)}.

\bibitem{Dudley2013}
Dudley, A.\ \textit{et al}.\ 
  Efficient sorting of Bessel beams.\ 
  \dois{10.1364/OE.21.000165}{\textit{Opt.\ Express} \textbf{21}, 165--171 (2013)}.
  
\bibitem{ferris}
Franke-Arnold, S. \textit{et al}.\ 
 Optical ferris wheel for ultracold atoms. \dois{10.1364/OE.15.008619}{\textit{Opt. Express} \textbf{15}, 8619 (2007).}
 
\bibitem{ferris2}
Arnold, A.~S. Extending dark optical trapping geometries.\  \dois{10.1364/OL.37.002505}{\textit{Opt.\ Lett.} \textbf{37}, 2505 (2012)}. 

 \bibitem{Offer2016}
Offer, R.~F., Conway, J.~W.~C., Riis, E., Franke-Arnold, S.\ \& Arnold, A.~S.\ Cavity-enhanced frequency up-conversion in rubidium vapor.\  
  \dois{10.1364/OL.41.002177}{\textit{Opt.\ Lett}.\ \textbf{41}, 2177--2180 (2016)}.

\bibitem{Kleinman1968}
 Boyd G.~D.\ \& Kleinman, D.~A.\
 Parametric interaction of focused Gaussian light beams.\  
  \dois{10.1063/1.1656831}{\textit{J.\ Appl.\ Phys}.\ \textbf{39}, 3597--3639 (1968)}.
  
  \bibitem{irina}
Zhang, M., Guidry, M.~A., Lanning, R.~N., Xiao, Z., Dowling, J.~P., Novikova, I. \& Mikhailov, E.~E.\ Multi-pass configuration for improved squeezed vacuum generation in hot Rb vapor.\ 
\dois{10.1103/PhysRevA.96.013835}{\textit{Phys.\ Rev.\ A} \textbf{96}, 013835 (2017)}.

\bibitem{ifan}
Zentile, M.~A., Whiting, D.~J., Keaveney, J.,  Adams, C.~S.\ \& Hughes, I.~G.\ Atomic Faraday filter with equivalent noise bandwidth less than 
$1\,$GHz.\ 
 \dois{10.1364/OL.40.002000}{\textit{Opt.\ Lett}.\  \textbf{40}, 2000--2003 (2015)}.

\bibitem{collective}
Peyrot, T.\ \textit{et al}. 
Collective Lamb shift of a nanoscale atomic vapor layer within a sapphire cavity.\ 
\dois{10.1103/PhysRevLett.120.243401}{\textit{Phys.\ Rev.\ Lett}.\ \textbf{120}, 243401 (2018)}.

\bibitem{rydberg}
Ripka, F., K\"{u}bler, H., L\"{o}w, R.\ \& Pfau, T.\ 
A room temperature single-photon source based on strongly interacting Rydberg atoms.
\href{https://arxiv.org/abs/1806.02120}{\textit{arxiv:1806.02120} (2018)}.

\bibitem{Nicolas2014}
Nicolas, A.\ \textit{et al}.\ 
  A quantum memory for orbital angular momentum photonic qubits.\ 
  \dois{10.1038/nphoton.2013.355}{\textit{Nature Phot}.\ \textbf{8}, 234--238 (2014)}.

\bibitem{Ding2015}
Ding, D.~S.\ \textit{et al}.\ 
 Quantum storage of orbital angular momentum entanglement in an atomic ensemble.\ 
  \dois{10.1103/PhysRevLett.114.050502}{\textit{Phys.\ Rev.\ Lett}.\ \textbf{114}, 050502 (2015)}.
 
\bibitem{Brekke2016}
Brekke E.\ \& Potier, S.\  
 Optical cavity for enhanced parametric four-wave mixing in rubidium.\  
  \dois{10.1364/AO.56.000046}{\textit{Appl.\ Opt}.\ \textbf{56}, 46--49 (2017)}.

\bibitem{Arnold1998}
Arnold, A.~S., Wilson, J.~S.\ \& Boshier, M.~G.\ 
 A simple extended-cavity diode laser.\ 
  \dois{10.1063/1.1148756}{\textit{Rev.\ Sci.\ Instrum}.\ \textbf{69}, 1236--1239 (1998)}.

\bibitem{Akulshin2012}
Akulshin, A., Perrella, C., Truong, G.-W., McLean, R.\ \& Luiten, A.\ 
 Frequency evaluation of collimated blue light generated by wave mixing in Rb vapour.\ 
 \dois{10.1088/0953-4075/45/24/245503}{\textit{J.\ Phys.\ B} \textbf{45}, 245503 (2012)}.

\bibitem{Clark2016}
Clark, T.~W., Offer, R.~F., Franke-Arnold, S., Arnold, A.~S.\ \& Radwell, N.\ 
  Comparison of beam generation techniques using a phase only spatial light modulator.\ 
  \dois{10.1364/OE.24.006249}{\textit{Opt.\ Express} \textbf{24}, 6249--6264 (2016)}.

\bibitem{Radwell2017}
Radwell, N., Offer, R.~F., Selyem, A.\ \& Franke-Arnold, S.\  
  Optimisation of arbitrary light beam generation with spatial light modulators.\ 
  \dois{10.1088/2040-8986/aa7f50}{\textit{J. Opt}.\ \textbf{19}, 095605 (2017)}.

\bibitem{Davis1999}
Davis, J.~A., Cottrell, D.~M., Campos, J., Yzuel, M.~J.\ \& Moreno, I.\ 
  Encoding amplitude information onto phase-only filters.\ 
  \dois{10.1364/AO.38.005004}{\textit{Appl.\ Opt}.\ \textbf{38}, 5004--5013 (1999)}.
  
 \bibitem{Offer2018}
Offer, R.~F., Stulga, D., Riis, E., Franke-Arnold, S.\ \& Arnold, A.~S.\ 
  Data for: Spiral bandwidth of four-wave mixing in Rb vapour.\ 
 \dois{10.15129/96db0ebb-aace-494f-8e61-d4a064fcadbb}{DOI: 10.15129/96db0ebb-aace-494f-8e61-d4a064fcadbb}.

}
\end{thebibliography}
\end{document}